\documentclass[11pt,fleqn]{article}
\usepackage{graphicx}
\usepackage{amsmath}

\usepackage{amsfonts,amssymb,cite}
\usepackage{epsf,graphicx}

\newcommand{\bls}[1]{\renewcommand{\baselinestretch}{#1}}

\def\noi{\noindent}
\def\jnumber#1#2{\thispagestyle{empty} \noi\unitlength=1mm
        \begin{picture}(178,10)
            \put(177,15){\llap{\large\it Grav. Cosmol. No.\,#1, #2}}
                    \end{picture}}

\newcommand{\Title}[1]{\noi {{\Large\bf #1}}\\[1ex]}

\def\Aunames#1{\noi{\bf #1}}

\def\Addresses#1{\medskip\noi \protect
    \begin{description}\itemsep -3pt {\it #1} \end{description}}
\def\addr#1#2{\item[${}^{#1}$]{\it #2}}

\newcommand{\Rec}[1]{\noi {\it Received #1} \\}

\newcommand{\Abstract}[1]{\vskip 2mm \begin{center}
        \parbox{16.4cm}{\small\noi #1} \end{center}\medskip}

\def\email#1#2{\footnotetext[#1]{e-mail: #2}\addtocounter{footnote}{1}}

\def\nqq{\hspace*{-2em}}
\def\nhq{\hspace*{-0.5em}}

\def\cm{\hspace*{1cm}}
\def\inch{\hspace*{1in}}

\def\Jl#1#2{#1 {\bf #2},\ }

\def\ApJ#1 {\Jl{Astroph. J.}{#1}}
\def\CQG#1 {\Jl{Class. Quantum Grav.}{#1}}
\def\DAN#1 {\Jl{Dokl. AN SSSR}{#1}}
\def\GC#1 {\Jl{Grav. Cosmol.}{#1}}
\def\GRG#1 {\Jl{Gen. Rel. Grav.}{#1}}
\def\JETF#1 {\Jl{Zh. Eksp. Teor. Fiz.}{#1}}
\def\JETP#1 {\Jl{Sov. Phys. JETP}{#1}}
\def\JHEP#1 {\Jl{JHEP}{#1}}
\def\JMP#1 {\Jl{J. Math. Phys.}{#1}}
\def\NPB#1 {\Jl{Nucl. Phys. B}{#1}}
\def\NP#1 {\Jl{Nucl. Phys.}{#1}}
\def\PLA#1 {\Jl{Phys. Lett. A}{#1}}
\def\PLB#1 {\Jl{Phys. Lett. B}{#1}}
\def\PRD#1 {\Jl{Phys. Rev. D}{#1}}
\def\PRL#1 {\Jl{Phys. Rev. Lett.}{#1}}

\def\al{&\nhq}
\def\lal{&&\nqq {}}

\def\eqs{Eqs.\,}
\def\beq{\begin{equation}}
\def\eeq{\end{equation}}
\def\bear{\begin{eqnarray}}
\def\bearr{\begin{eqnarray} \lal}
\def\ear{\end{eqnarray}}
\def\earn{\nonumber \end{eqnarray}}

\def\nnn{\nonumber\\ \lal }

\def\yyy{\\[5pt] \lal }
\def\eql{\al =\al}

\def\sequ#1{\setcounter{equation}{#1}}

\def\d{\partial}

\def\diag{\mathop{\rm diag}\nolimits}

\def\const{{\rm const}}

\bls{1.025}
\begin{document}
\twocolumn[
\jnumber{2}{2011}

\Title{Rotating thin-shell wormhole from glued Kerr spacetimes}

\Aunames{P. E. Kashargin$^a$ and S. V. Sushkov$^{a,b,1}$}

\Addresses{
\addr a {Department of General Relativity and Gravitation,
    Kazan State University, Kremlevskaya str. 18, Kazan 420008,
    Russia}
\addr b {Department of Mathematics, Tatar State University of Humanities and
    Education, Tatarstan str. 2, Kazan 420021, Russia}
      }

\Rec{January 13, 2011}

\Abstract
 {We construct a model of a rotating wormhole made by cutting and pasting
  two Kerr spacetimes. As a result, we obtain a rotating thin-shell wormhole
  with exotic matter at the throat. Two candidates for the exotic matter are
  considered: (i) a perfect fluid; (ii) an anisotropic fluid. We show that
  a perfect fluid is unable to support a rotating thin-shall wormhole.
  On the contrary, the anisotropic fluid with the negative energy density
  can be a source for such a geometry.
  }

\bigskip

] 
\email 1 {sergey\_sushkov@mail.ru; sergey.sushkov@ksu.ru}

{ 
\newcommand{\Ref}[1]{(\ref{#1})}
\newcommand{\hi}{{\hat\imath}}
\newcommand{\hj}{{\hat\jmath}}
\newcommand{\cE}{{\cal E}}
\newcommand{\cP}{{\cal P}}
\def\ee{{\mathbf{e}}}

\section{Introduction}

  Wormholes are usually defined as topological handles in spacetime linking
  widely separated regions of a single universe, or ``bridges'' joining two
  different spacetimes [1]. Their history traces back to the works of
  Einstein and Rosen [2], and Misner and Wheeler [3]. The modern interest in
  wormholes dates back to 1988, when Morris and Thorne [4] discussed the
  possibility of using wormholes for interstellar travels. As is well-known
  [4, 5], traversable wormholes can exist only if their throats contain
  exotic matter which possesses a negative pressure and violates the null
  energy condition. The search for realistic physical models providing the
  wormhole existence represents an important direction in wormhole physics.
  In general relativity there are models of wormholes supported by matter
  with exotic equations of state such as phantom energy [6, 7], a Chaplygin
  gas [8], tachyon matter [9]. Numerous examples of wormhole solutions have
  been found in various modifications of general relativity such as
  scalar-tensor theories of gravity, brane theories, semiclassical gravity,
  theories with non-minimal coupling [10, 11]. It is worth being noticed
  that most of the investigations deal with static spherically symmetric
  wormholes because of their simplicity and high symmetry.  At the same
  time, it would be important and interesting from a physical point of view
  to study wider classes of wormholes including non-static and rotating ones.

  Rotating wormholes were first considered by Teo [12] who discussed some
  general geometrical properties of the stationary rotating wormhole
  spacetime. Other investigations in this field include studies of general
  requirements to the stress-energy tensor necessary to generate a rotating
  wormhole [13], energy conditions in a rotating wormhole spacetime and its
  traversability [14], and scalar perturbations in the rotating wormhole
  background [15]. Arguments in favor of the possibility of existence
  of semiclassical rotating wormholes were given in [16]. Solutions
  describing slowly rotating wormholes have been found and analyzed in [17,
  18]. A number of new axially symmetric stationary exact solutions in
  general relativity with phantom and Maxwell fields have recently been
  obtained in [19, 20]; among them are solutions which represent rotating
  and magnetized wormholes.

  The first examples of thin-shell wormholes have been given by Visser [21,
  22]. In particular, he considered a spherically symmetric thin-shell
  wormhole constructed by joining two Schwarzschild geometries [22].
  Generally, thin-shell wormholes are made by cutting and pasting two
  manifolds to form a geodesically complete new one with a throat located on
  the joining shell. In this case, the exotic matter needed to build the
  wormhole is concentrated on the shell, and the junction-condition
  formalism is used for its study. Due to elegancy and relative simplicity,
  the cut-and-paste approach has became generally used for constructing new
  models of thin-shell wormholes such as charged wormholes [23], those with
  a cosmological constant [24], cylindrical [25] and plane-symmetric [26]
  wormholes, those in dilaton [27], Einstein--Gauss--Bonnet [28, 29], and
  Brans-Dicke [30] gravity, wormholes with a generalized Chaplygin gas [31],
  wormholes associated with global cosmic strings [32] and global monopoles
  [33]. Worth mentioning is also the paper by Bronnikov and Starobinsky [34]
  who considered static, spherically symmetric thin-shell wormholes in any
  non-ghost scalar-tensor theory of gravity and showed that the shell
  surface energy density is negative in all such cases.

  In this paper we will apply the cut-and-paste method in order to construct
  and study a rotating thin-shell wormhole made by cutting and pasting two
  Kerr spacetimes.

\section{Kerr surgery}

  The Kerr metric in the Boyer-Lindquist coordinates reads [35]
\bearr
    ds^2=\left(1-\frac{2mr}{\rho^2}\right)dt^2
        -\frac{\rho^2}{\Delta}dr^2-\rho^2d\theta^2
\nnn \cm
    -\left(r^2+a^2+\frac{2ma^2r}{\rho^2}\sin^2\theta\right)\sin^2\theta
    d\phi^2
\nnn \cm
    +\frac{4mar}{\rho^2}\sin^2\theta d\phi dt, \label{kash-metric}
\ear
  where $\rho^2 = r^2+a^2\cos^{2}\theta$ and $\Delta = r^2-2mr +a^2$.
  The parameters $m$ and $J=ma$ correspond to the mass and angular momentum
  of a Kerr black hole measured by a distant observer. The metric (1)
  has two fictitious singularities. The first one occurs at the {\em event
  horizon} $r=r_{+}$ where $\Delta=0$, and hence $g_{rr}$ is infinite:
\beq \label{kash-r+}
    r_+ = m + \sqrt{m^2-a^2}.
\eeq
  The second singularity occurs on the boundary of the {\em ergosphere}
  $r=r_0$ where $g_{tt}=0$:
\beq \label{kash-r_0}
    r_0 = m+\sqrt{m^2-a^2\cos^2\theta}.
\eeq

  Consider two copies ${\cal M}_1$ and ${\cal M}_2$ of the region
  $r\ge b$ of the Kerr spacetime (\ref{kash-metric}):
\beq
    {\cal M}_{1,2}=\{ (t,r,\theta,\phi) \, |\, r\ge b \}.
\eeq
  As a result, we get two geodesically incomplete manifolds with boundaries
  given by the timelike hypersurfaces
\beq
    \Sigma_{1,2}=\{ (t,r,\theta,\phi) \, |\, F(r)=r-b=0 \}.
\eeq
  Identifying these hypersurfaces (i.e., $\Sigma = \Sigma_{1} \equiv
  \Sigma_{2}$), we obtain a new manifold ${\cal M}={\cal M}_{1}\cup{\cal
  M}_{2}$, which is geodesically complete and possesses two asymptotically
  flat regions connected by a wormhole with the throat $\Sigma$. Note that
  the two-dimen\-sional surface $t=\const$, $r=b$ in Kerr spacetime is
  actually an ellipsoid of revolution having minor and major axes equal to
  $b$ and $2(a^2+b^2)^{1/2}$, respectively. Nevertheless, for brevity we
  will call $b$ the wormhole throat radius. To avoid the presence of
  horizons in the resulting manifold $\cal M$, we will suppose $ b > r_{+}$.
  Since $\cal M$ is piecewise Kerr, the stress-energy tensor is everywhere
  zero, except for the throat itself. At $\Sigma$ one may expect a
  stress-energy tensor proportional to the delta function. This means that
  the throat $\Sigma$ is a thin shell.

  To analyze such a thin-shell configuration, we will follow the
  Darmois-Israel standard formalism [36], also known as the junction
  condition formalism. The wormhole throat $\Sigma$ is a synchronous
  timelike hypersurface, where we define the intrinsic coordinates
  $\xi^{i} = (\tau,\vartheta,\varphi)$ as follows:  $\tau=t_1\equiv t_2$,
  $\vartheta=\theta_1\equiv \pi-\theta_2$, and $\varphi=\phi_1\equiv\phi_2$.
  The coordinate $\tau$ is the proper time on the shell. Generally, the
  throat radius can be a function of proper time. However, we will assume
  $b(\tau)\equiv b=\const$. Note that the metric (the first fundamental
  form) is continuous on $\Sigma$:
\beq
    g_{ij}^{1}|_\Sigma=g_{ij}^{2}|_\Sigma,\label{kash-g_on_sigma}
\eeq
  while its first derivatives can be discontinuous. To describe this
  discontinuity, one should consider the extrinsic curvature. The
  extrinsic curvatures (second fundamental forms) associated with the two
  sides of the shell $\Sigma$ are
\beq
    K^{\pm}_{ij}= \left. -n^\pm_{\gamma}\left( \frac{\d^2
    x^{\gamma}}{\d\xi^{i}\d\xi^{j}}+\Gamma^\gamma_{\alpha\beta}
    \frac{\d x^{\alpha}}{\d\xi^{i}} \frac{\d x^{\beta}}
    {\d\xi^{j}}\right)\right|_{\Sigma},
            \label{kash-second_form}
\eeq
  where $n^\pm_\gamma$ are the unit normals ($n^\gamma n_\gamma=1$)
  to $\Sigma$:
\beq
    n^\pm_\gamma=\pm\left|g^{\alpha\beta}\frac{\d F}{\d x^\alpha}
    \frac{\d F}{\d x^\beta}\right|^{-1/2}\frac{\d F}{\d x^\gamma}.
\eeq
  Generally, $K_{ij}^{+}\neq K_{ij}^{-}$. With the definitions
  $k_{ij} = K^{+}_{ij}-K^{-}_{ij}$ and $k=k^i_i$ we have the Einstein
  equations on the shell (also called the Lanczos equations)
\bear    \label{kash-Lanczos}
    -k_{ij} + kg_{ij} = 8\pi S_{ij},
\ear
  where $S_{ij}$ is the surface stress-energy tensor.

  Let us adopt the orthonormal basis
  $\{\ee_{\hat\tau},\ee_{\hat\vartheta},\ee_{\hat\varphi}\}$
  for the metric \Ref{kash-metric} on $\Sigma$:
\bearr  \label{kash-orthobasis}
    \ee_{\hat\tau} = \frac{\ee_{\tau}-\frac{g_{\tau\varphi}}
    {g_{\varphi\varphi}}  \ee_{\varphi}}
    {\sqrt{g_{\tau\tau}-\frac{g_{\tau\varphi}^2}{g_{\varphi\varphi}}}},
\nnn
    \ee_{\hat\vartheta} = \frac{\ee_{\vartheta}}
        {\sqrt{-g_{\vartheta\vartheta}}},
\nnn
    \ee_{\hat\varphi} = \frac{\ee_{\varphi}}{\sqrt{-g_{\varphi\varphi}}}.
\ear
  In this basis, the surface stress-energy tensor $S_{ij}$ has the
  following algebraic structure:
\beq    \label{kash-Sgen}
    S_{\hat\imath\hat\jmath} = \left[
    \begin{array}{ccc}
        \sigma  & 0                 & \zeta  \\
        0       & p_{\vartheta} & 0  \\
        \zeta     & 0                 & p_{\varphi}
    \end{array} \right],
\eeq
  where $\sigma$ is the surface energy density, $p_{\vartheta}$ and
  $p_{\varphi}$ are the principal surface pressures, and $\zeta$ is the
  surface angular momentum density. The Lanczos equations \Ref{kash-Lanczos}
  in the basis \Ref{kash-orthobasis} take the following form:
\begin{subequations}                            \label{kash-alleqs}
\bearr
    4\pi\sigma =  -\frac{\Delta_\beta^{1/2}}{m \rho_\beta
    \Phi}\Big[2\beta^3+\alpha^2\beta+\alpha^2
\nnn \inch
    +\alpha^2(\beta-1)\cos^2\vartheta\Big] , \label{kash-sigma}
\yyy
    4\pi p_{\vartheta} = \frac{\beta-1}{m\rho_\beta\Delta_\beta^{1/2}},
\yyy
    4\pi p_{\varphi} = \frac 1 {m \rho_\beta^3\Delta_\beta^{1/2}\Phi}
    \Big[\beta^2 \big(\beta^5-\beta^4+2\alpha^2\beta^3
\nnn \cm
    +2\alpha^2\beta^2 +\alpha^2\beta(\alpha^2-8)+3\alpha^4\big)
\nnn \cm
    +\alpha^2\cos^2\vartheta
    \big(\beta^5-5\beta^4+2\beta^3(\alpha^2+4)
\nnn \cm
    -6\alpha^2\beta^2 +\alpha^4\beta-\alpha^4\big)\Big] ,\label{kash-pphi}
\yyy
    4\pi \zeta = -\frac1{m\rho_\beta^3 \Phi}
    \Big[\alpha \sin\vartheta \big(3\beta^4+\alpha^2\beta^2
\nnn \cm\cm
    + \alpha^2(\beta^2-\alpha^2)\cos^2\vartheta \big)\Big],
        \label{kash-zeta}
\ear
\end{subequations}
  where we have introduced the convenient dimensionless quantities
$$
    \beta = bm^{-1},\qquad \alpha = am^{-1},
$$   $$
    \Delta_\beta = \beta^2-2\beta +\alpha^2, \qquad
    \rho_\beta^2 = \beta^2+\alpha^2\cos^{2}\theta,
$$   $$
    \Phi = \beta^4+\alpha^2\beta^2+2\alpha^2\beta
        +\alpha^2\Delta_\beta\cos^2\vartheta.
$$
{ Later on we will also use dimensionless notations for the event
horizon $\beta_+=r_+m^{-1}=1+\sqrt{1-\alpha^2}$ and the boundary
of ergosphere
$\beta_0=r_0m^{-1}=1+\sqrt{1-\alpha^2\cos^2\theta}$.}

\section{Matter on the shell}

  It is necessary to emphasize that the quantities $\sigma$,
  $p_{\vartheta}$, $p_{\varphi}$, and $\zeta$ given by \eqs (12) are not yet
  related to any physical model of matter filling the shell $\Sigma$. Their
  values are of purely geometric nature and depend on the metric parameters
  $m$ and $a$ and the throat radius $b$. To impart a physical sense to these
  quantities one should specify the kind of matter which can support the
  rotating thin-shell wormhole.

\subsection{Perfect fluid}

  As a simple model of matter located on the shell $\Sigma$, we will first
  consider a perfect fluid. In the orthonormal basis \Ref{kash-orthobasis}
  the surface stress-energy tensor of a perfect fluid is
\beq \label{kash-Sperfluid}
    S_{\hi\hj} = (\cE+\cP) u_\hi u_\hj - \eta_{\hi\hj}\cP,
\eeq
  where $\eta_{\hi\hj} = \diag(+1,-1,-1)$, $u_\hi$ is the fluid velocity
  which is supposed to be timelike, i.e. $u^\hi u_\hi = 1$, ${\cE}$ is the
  fluid energy density measured in the comoving frame, and $\cP$ is the
  pressure isotropic in all directions tangent to the shell $\Sigma$. For
  the rotating fluid it is naturally to choose $u_\hi= (u_\tau,0,u_\varphi)$.
  Comparing \Ref{kash-Sgen} and \Ref{kash-Sperfluid}, we find
\begin{subequations}                \label{kash-alleqs0}
\bearr
    \sigma = ({\cE}+\cP)u_\tau^2-\cP,
\\ \lal
    p_\vartheta  = \cP,
\\ \lal
    p_\varphi = ({\cE}+\cP)u_{\varphi}^2+\cP,
\\  \lal
    \zeta = ({\cE}+\cP)u_\tau u_\varphi.
\ear
\end{subequations}
  Combining these equations, one can easily obtain the following relation:
\beq                \label{kash-relperfluid}
    (\sigma+p_\vartheta)(p_\varphi-p_\vartheta)-\zeta^2\equiv 0.
\eeq
  Substitution of \eqs \Ref{kash-alleqs} into the last relation gives
\beq
    4\alpha^2\beta^2\rho_\beta^{-6}\sin^2\theta\equiv 0.
\eeq
  This identity is only fulfilled provided $\alpha = am^{-1} = 0$, i.e.,
  $a = 0$. Therefore, a perfect fluid cannot be a source for a rotating
  thin-shell wormhole with $a \ne 0$.\footnote
    {Nevertheless, a perfect fluid can support a spherically symmetric
    thin-shell wormhole made of two surgically modified Schwarzschild
    spacetimes [22].}

\subsection{Anisotropic fluid}

  Now consider an anisotropic fluid with the surface stress-energy tensor
\beq \label{kash-Sfluid}
    S_{\hi\hj} = \cE u_\hi u_\hj+\cP_1 v_\hi v_\hj+\cP_2 \Pi_{\hi\hj},
\eeq
  Here $u_\hi  =  (u_\tau,0,u_\varphi)$ is the fluid timelike velocity
  ($u^\hi u_\hi = 1$), and $v_\hi$ and $\Pi_{\hi\hj}$ satisfy the following
  orthogonality conditions:
\beq
    u^\hi v_\hi = 0, \quad u^\hi\Pi_{\hi\hj} = 0, \quad
        v^\hi\Pi_{\hi\hj} = 0.
\eeq
  $\cE$ is the energy density, $\cP_1$ and $\cP_2$ are the fluid pressures
  in two orthogonal directions tangent to the shell $\Sigma$ (generally,
  $\cP_1\not = \cP_2$). For the rotating fluid it is natural to choose
  $u_\hi  =  (u_\tau,0,u_\varphi)$ with
\beq \label{kash-norm}
    u_\tau^2-u_\varphi^2 = 1,
\eeq
  and $v_\hi = (0,1,0)$; the tensor $\Pi_{\hi\hj}$ can be constructed
  as follows: $\Pi_{\hi\hj} = u_\hi u_\hj-v_\hi v_\hj-\eta_{\hi\hj}$.
  Comparing \Ref{kash-Sgen} and \Ref{kash-Sfluid}, we find
\begin{subequations}                          \label{kash-alleqs2}
\bearr
    \sigma = ({\cE}+\cP_2)u_\tau^2-\cP_2 ,
\\ \lal
    p_\vartheta = \cP_1,
\\ \lal
    p_\varphi = ({\cE}+\cP_2)u_{\varphi}^2+\cP_2,
\\ \lal
    \zeta  =  ({\cE}+\cP_2)u_\tau u_\varphi,
\ear
\end{subequations}
  The latter equations, together with the normalizing condition
  \Ref{kash-norm}, form a set of five algebraic equations for five unknowns
  $\cE$, $\cP_1$, $\cP_2$, $u_\tau$ and $u_\varphi$.  Resolving the system
  yields $\cP_1 =  p_\vartheta$, and
\begin{subequations}                        \label{kash-sol}
\bear
    \cE^\pm \eql \frac12\left[ \sigma-p_\varphi\pm\sqrt{D}\right],
        \label{kash-rho_pm}
\\
    \cP_2^\pm \eql \frac12\left[-\sigma+p_\varphi\pm\sqrt{D}\right],
        \label{kash-P_p_pm}\\
    (u^\pm_\tau)^2 \eql \pm\frac{\sigma+p_\varphi}{2\sqrt{D}}+\frac12,
    \label{kash-ut_pm}
\\
    (u^\pm_\varphi)^2 \eql
    \pm\frac{\sigma+p_\varphi}{2\sqrt{D}}-\frac12, \label{kash-up_pm}
\ear
\end{subequations}
  with $D = (\sigma+p_\varphi)^2-4\zeta^2$. It is worth noting that we have
  got two classes of solutions which depend on a choice of the plus or minus
  sign in the obtained expressions.

  Finally, \eqs \Ref{kash-sol} represent expressions for the surface energy
  density $\cE$, pressures $\cP_1$ and $\cP_2$, and velocity components
  $u_\tau$ and $u_\varphi$ of the anisotropic fluid on the shell $\Sigma$.

\section{Analysis}

  In this section we will analyze the model of a rotating thin-shell
  wormhole constructed above. First of all, let us consider the particular
  case of a non-rotating thin-shell wormhole with $a = 0$ (no angular
  momentum).  In this case the metric \Ref{kash-metric} reduces to the
  Schwarzschild one, and Eqs. \Ref{kash-alleqs} reduce to those obtained by
  Visser [22]:
\bearr
        \sigma = -\frac{1}{2\pi b}\sqrt{1-2m/b},
\nnn
    p_\vartheta = p_\varphi = \frac{1}{4\pi b}\frac{1-m/b}{\sqrt{1-2m/b}},
    \qquad \zeta = 0.
\ear
  Note that the surface energy density $\sigma$ tends to zero and the
  pressures $p_\vartheta$ and $p_\varphi$ to infinity if the throat radius
  $b$ tends to that of the event horizon $r_g = 2m$.

  In the general case of a rotating thin-shell wormhole with $a\not = 0$
  we have $\sigma \sim \Delta_\beta^{1/2}$ and $p_\vartheta,\
  p_\varphi\sim \Delta_\beta^{-1/2}$ (see \Ref{kash-alleqs}). Since
  $\Delta_\beta = 0$ if $\beta = \beta_+: = 1+\sqrt{1-\alpha^2}$, we can
  see that $\sigma\to 0$ and $p_\vartheta,\ p_\varphi \to\infty$
  as $\beta\to \beta_+$.

  Now let us discuss the properties of the anisotropic fluid located on
  the shell $\Sigma$. Given the expressions \Ref{kash-alleqs} for $\sigma$,
  $p_\vartheta$, $p_\varphi$, and $\zeta$, we can find the values $\cE$,
  $\cP_1$, $\cP_2$, $u_\tau$, and $u_\varphi$ as explicit functions of the
  dimensionless throat radius $\beta$. In particular, we have
\bearr
    D = \frac1{4\pi m\rho_\beta^{6}\Delta_\beta}
    \Big[\beta^3(\beta(\beta-3)^2-4\alpha^2)
\nnn
    +2\alpha^2\beta\cos^2\vartheta
    (\beta^3-3\beta+2\alpha^2)+\alpha^4\cos^4\vartheta(\beta-1)^2\Big].
\nnn
\ear
  Note that $D$ should necessarily be positive, i.e., $D > 0$. As is shown
  in the Appendix, it is possible if and only if $\beta\in{I}_1\cup{I}_2$,
  where ${I}_1 = (\beta_+,\beta_2)$, ${I}_2 = (\beta_3,\infty)$, and
\beq            \label{kash-beta_n}
    \beta_n = 2+2\cos\left(\frac{\chi-2\pi(3-n)}{3}\right), \quad
            n = 1,2,3,
\eeq
  with $\chi$ defined by $\cos\chi = 2\alpha^2-1$. Additionally, one should
  check whether or not the values of $(u_\tau^\pm)^2$ and
  $(u_\varphi^\pm)^2$ given by \eqs \Ref{kash-ut_pm} and \Ref{kash-up_pm}
  are non-negative.\footnote
    {In principle, one may discard this requirement and consider also
    negative values of $u_\tau^2$ and $u_\varphi^2$. In this case the
    components $u_\tau$ and $u_\varphi$ will be pure imaginary, and
    as a consequence $u^\hi$ will be spacelike, i.e. $u^\hi u_\hi = -1$.
    In turn, this means that the fluid velocity exceeds the
    velocity of light.}
  From Fig.\,1 one may see that $(u_\tau^+)^2$ and $(u_\varphi^+)^2$ are
  positive if $\beta<\beta_2$, while $(u_\tau^-)^2$ and $(u_\varphi^-)^2$
  are positive if $\beta>\beta_3$. This means that one should take the plus
  sign in \eqs \Ref{kash-rho_pm}--\Ref{kash-up_pm} in the case $\beta\in
  I_1$ and the minus sign if $\beta\in I_2$. Let us repeat that the domain
  $\beta \le \beta_+$ is forbidden by definition since we consider only
  wormholes whose throat radius is greater than that of the event horizon
  $\beta_+$. In addition, it turns out that the domain
  $\beta\in[\beta_2,\beta_3]$ is also forbidden for rotating thin-shell
  wormholes. Thus we have two classes of wormhole solutions depending on
  the throat radius $\beta$: (i) $\beta_+ < \beta < \beta_2$; (ii)
  $\beta > \beta_3$.

  The energy density $\cE$ and the pressures $\cP_1$ and $\cP_2$ as
  functions of $\beta$ are shown in Fig.\,2. Note that $\cE$ is negative,
  while $\cP_1$ and $\cP_2$ are positive for all values of $\beta$.

\begin{figure}[ht]
\centerline{\includegraphics[scale = 0.4]{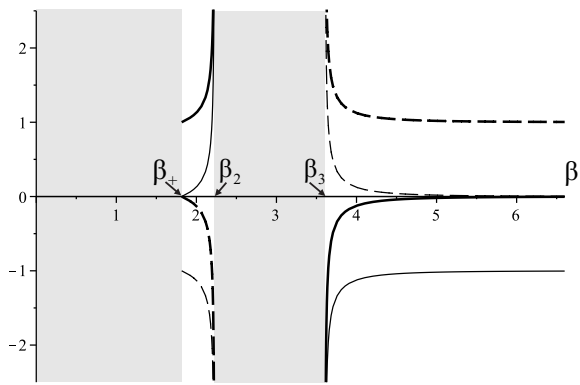}}
\caption{Plots of $(u_\tau^\pm)^2$ and $(u_\varphi^\pm)^2$ vs. $\beta$ with
    given $\alpha = 0.5$, $m = (4\pi)^{-1}$. The solid and dashed curves
    are used for the plus- and minus-sign solutions, respectively; thick
    lines show $(u_\tau^\pm)^2$, and thin lines show
    $(u_\varphi^\pm)^2$. The shaded areas indicate forbidden regions
    $\beta \le \beta_+$ and $\beta\in[\beta_2,\beta_3]$.
\label{kash-fig1}}
\end{figure}

\begin{figure}[ht]
\centerline{\includegraphics[scale = 0.4]{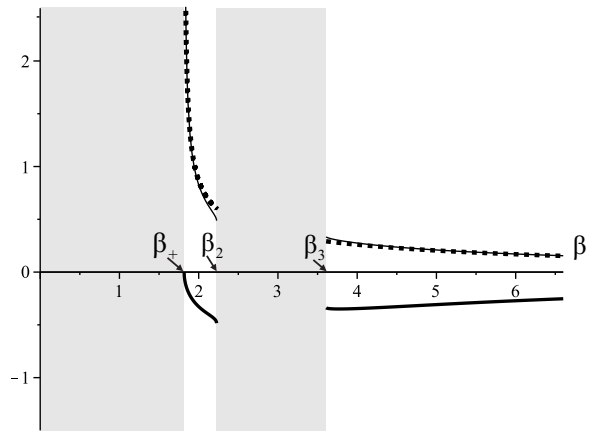}}
\caption{Plots of $\cE$, $\cP_1$, and $\cP_2$ vs. $\beta$ with given $\alpha
     = 0.5$, $m = (4\pi)^{-1}$. Solid, dotted, and thick lines
     show $\cE$, $\cP_1$, and $\cP_2$, respectively. The shaded areas
     indicate forbidden regions $\beta \le \beta_+$ and
$\beta\in[\beta_2,\beta_3]$. \label{kash-fig2}}
\end{figure}

\section{Conclusion}

  We have constructed a rotating wormhole model by cutting and pasting
  two Kerr spacetimes. As is usual for the cut-and-paste approach, the
  resulting wormhole spacetime has a thin shall joining two regions of Kerr
  spacetimes. This shell represents the wormhole throat and contains exotic
  matter needed to support the wormhole. We have discussed two possible
  candidates to the role of the exotic matter: (i) a perfect fluid, and (ii)
  an anisotropic fluid. It has been shown that a perfect fluid is unable
  to support a rotating thin-shell wormhole, while an anisotropic fluid
  localized on the shell can be a source of such geometry. The corresponding
  fluid energy density $\cE$ and anisotropic pressures $\cP_1$ and $\cP_2$
  are given by \eqs \Ref{kash-sol} which express $\cE$, $\cP_1$, and
  $\cP_2$ as functions of the dimensionless throat radius $\beta$.
{Admissible values of $\beta$ belong to two nonintersecting
intervals $I_1=(\beta_+,\beta_2)$ and $I_2=(\beta_3,\infty)$,
where $\beta_+=1+\sqrt{1-\alpha^2}$ is the event horizon, and
$\beta_n$ ($n=2,3$) are given by Eq. \Ref{kash-beta_n}. Since
$\beta_2<\beta_3$, the intervals $I_1$ and $I_2$ are not
intersected. Therefore, }
  there are two classes of wormhole solutions: (i) with
  ``small'' throat radii $\beta_+<\beta<\beta_2$, and (ii) with
  ``large'' radii $\beta>\beta_3$. In both cases the energy
  density $\cE$ of the anisotropic fluid turns out to be negative.
  This means that matter supporting the rotating wormhole violates
  the weak energy condition.

{It is interesting that the throat radius $\beta$ of the rotating
thin-shell wormhole can be less than the maximal size of
ergosphere $\beta_0^{max}=2$ ($\theta=\pi/2$). This is possible
for wormholes of the class I with small throat radii
$\beta_+<\beta<\beta_2$ (see the appendix). Moreover, for
wormholes with large angular momentum $\alpha>2^{-1/2}$ all values
of $\beta$ from the interval $(\beta_+,\beta_2)$ are less than
$\beta_2^{max}$. Thus, there are wormholes (of the class I) whose
throat lies inside of the ergosphere. Such the feature may, in
principle, lead to interesting consequences due to processes
similar to the Penrose process in the ergosphere of Kerr black
hole.}

  An important issue in wormhole physics is the stability of wormhole
  configurations. The stability of spherically symmetric thin-shell
  wormholes has been intensively considered in the literature [37--44].
  We intend to study this problem for rotating thin-shall wormholes in our
  forthcoming paper.

\section*{Appendix}
\def\theequation{A.\arabic{equation}}
\sequ 0

  Rearranging \eqs \Ref{kash-alleqs2} yields
\beq
    \zeta^2  =  (\sigma+\cP_2)(p_\varphi-\cP_2) , \label{kash-Pphi}
\eeq
  It is a quadratic equation for $\cP_2$ with the discriminant $D =
  (\sigma+p_\varphi)^2-4\zeta^2$ which should be necessarily positive,
  $D > 0$. Using the relations \Ref{kash-sigma}, \Ref{kash-pphi}, and
  \Ref{kash-zeta}, we find
\bearr              \label{kash-D}
    D = (4\pi m)^{-1}\rho_0^{-6}\Delta_0^{-1}
    \Big[\beta^3(\beta(\beta-3)^2-4\alpha^2)
\nnn  \cm
    +2\alpha^2\beta\cos^2\vartheta  (\beta^3-3\beta+2\alpha^2)
\nnn  \cm
    +\alpha^4\cos^4\vartheta (\beta-1)^2\Big].
\ear
  Since $b > r_{+}$ is assumed, we have $\beta>\beta_+ = 1+\sqrt{1-\alpha^2}$,
  and one may check in a straightforward manner that the cosine terms in
  \Ref{kash-D} are positive.  Therefore the sign of $D$ is determined by the
  first term in the square brackets. In particular, on the equator
  $\vartheta = \pi/2$ the condition $D>0$ reduces to
\beq    \label{kash-ineqq}
    f_\alpha(\beta) = \beta(\beta-3)^2-4\alpha^2 > 0.
\eeq
  The cubic parabola $f_\alpha(\beta)$ has three roots $\beta_n$
  ($n = 1,2,3$) given by Cardano's formulas:
\beq
    \beta_n = 2+2\cos\left(\frac{\chi-2\pi(3-n)}{3}\right),
\eeq
  with $\chi$ defined by
\[
    \cos\chi = 2\alpha^2-1.
\]
  In the case $0 < \alpha  < 1$ all roots are real and different, such
  that $\beta_1 < \beta_2 < \beta_3$; if $\alpha = 0$, then $\beta_1 = 0$
  and $\beta_2 = \beta_3 = 3$; if $\alpha = 1$, then $\beta_1 = \beta_2 = 1$
  and $\beta_3 = 4$
  {(see Fig. \ref{fig3})}.
  Formally, one can also consider $\alpha > 1$ (i.e.,
  $a > m$); in this case $\beta_1$ and $\beta_2$ become imaginary, and
  $\beta_3$ is an only real root. In general, the solution of the inequality
  \Ref{kash-ineqq} reads
\[
     \beta\in(\beta_1,\beta_2)\cup(\beta_3,\infty).
\]
  In addition, let us recall that it is assumed $b > r_{+}$,  hence
  $\beta > \beta_{+} = 1 + \sqrt{1-\alpha^2}$. One can check that
  $\beta_1 < \beta_+ < \beta_2$, and so we finally have
\beq\label{kash-interval}
    \beta \in (\beta_+, \beta_2) \cup (\beta_3,\infty).
\eeq

{
\begin{figure}[ht]
\centerline{\includegraphics[scale=0.35]{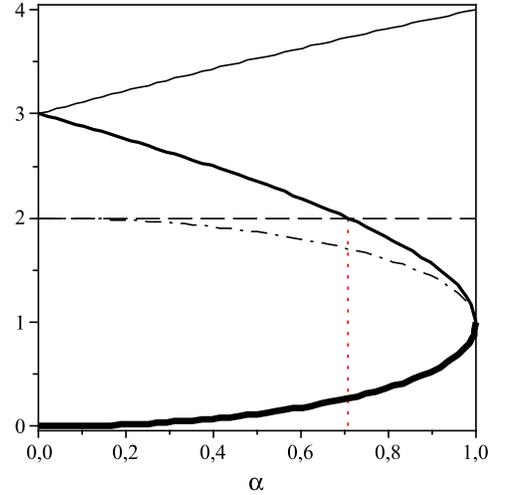}}
\caption{Graphs of roots $\beta_n$ vs. $\alpha$. Thick, middle,
and thin lines denote $\beta_1$, $\beta_2$, and $\beta_3$,
respectively. The dot-dashed line indicates the event horizon
$\beta_+=1+\sqrt{1-\alpha^2}$. The dashed line shows the maximal
size of ergosphere $\beta_0^{max}=2$ ($\theta=\pi/2$). The lines
for $\beta_2$ and $\beta_0^{max}$ are intersected at
$\alpha=2^{-1/2}$. \label{fig3}}
\end{figure}
Thus, admissible values of $\beta$ belong to two nonintersecting
intervals $I_1=(\beta_+,\beta_2)$ and $I_2=(\beta_3,\infty)$. Note
that they can only be intersected if $\alpha=0$ (no rotation),
when $\beta_2=\beta_2=3$. In this case one may obtain static,
spherically symmetric thin-shell wormhole with the throat's radius
$\beta=3$, or $b=3m$, whose value lies on the boundary between
$I_1$ and $I_2$ \cite{kash-Vis89b}.

It is also worth emphasizing that an admissible value of $\beta$
can be less than the maximal size of ergosphere $\beta_0^{max}=2$
($\theta=\pi/2$). Really, in case $\beta\in(\beta_+,\beta_2)$ one
may always choose $\beta_+<\beta<\min(\beta_0^{max},\beta_2)$ (see
Fig. \ref{fig3}). Moreover, for $\alpha>2^{-1/2}$ one has
$\beta_2<2$, hence all values of $\beta$ from the interval
$(\beta_+,\beta_2)$ are less than $\beta_2^{max}$.

}

\subsection*{Acknowledgments}

  The authors are deeply grateful to Kirill Bronnikov for a valuable
  discussion.
  The work was partially supported by the Russian Foundation for Basic
  Research grants No 08-02-00325, 08-02-91307.

} 

\end{document}